\documentclass{optica-article}

\journal{opticajournal} 

\articletype{Research Article}

\usepackage{lineno}
\usepackage{booktabs}
\usepackage{amsmath}

\begin{document}

\title{Recovery of phase constant from two-photon interference pattern by phase retrieval algorithm}

\author{
Yuhang Lei,\authormark{1} 
Wen Zhao,\authormark{1} 
Liang Cui,\authormark{1,*} 
and Xiaoying Li\authormark{1}}

\address{
\authormark{1}College of Precision Instrument and Opto-Electronics Engineering, Key Laboratory of Opto-Electronics
Information Technology, Ministry of Education, Tianjin University, Tianjin 300072, China}

\email{\authormark{*}Lcui@tju.edu.cn
} 


\begin{abstract*} 
For a HOM interferometer with two independent incident pulses, the interference pattern can be affected by adding a dispersion medium on one of the incident directions, 
but there hasn’t been a method to reconstruct the phase constant of the medium from the interference pattern.
To solve it, we adapted two phase retrieval algorithms and used them to recover the phase difference function between the two incident fields, from which the phase constant can be derived.
Through simulations, we verified the convergence, accuracy, and robustness of the algorithms, indicating that this phase recovery process can be completed well with negligible error.
Our research finds a new application direction for the phase recovery algorithm, provides an algorithmic tool for high-order dispersion measurement using two-photon interference, and paves the way for a higher resolution and phase-sensitive quantum tomography.
\end{abstract*}

\section{Introduction}

Hong-Ou-Mandel(HOM) interference is the bunching effect when identical fields incident a beam splitter(BS) from different directions. The HOM effect is depicted by the reduction of coincidence rate, which is also known as a HOM dip, and it can be affected by identicalness between incident fields, 
including their polarization, quantum states, position, etc \cite{PhysRevLett.59.2044, PhysRevResearch.4.023125, PhysRevLett.96.240502}. 

For HOM interference with two independent pulses, the visibility function, which can be derived from the HOM dip is related to the states and their cross-correlation function of the incident pulses. 
As the cross-correlation function is also the Fourier transform of the cross-spectrum density function which contains incident pulses’ phase information, introducing dispersion to the incident fields will have a predictable effect on the HOM interference patterns \cite{2008Observation, Xiaoxin2015Hong,ma2011effect,fan2021effect}.

Conversely, the HOM effect might be used to measure the dispersion of mediums.
Wen Zhao et al. have built an unbalanced Mach-Zender interferometer using the HOM effect and clarified that the group velocity dispersion(GVD) of the medium in the signal arm can be derived through the measured visibility function when incident pulses are in specific modes \cite{PhysRevResearch.4.023125, 2021Propagation,fan2021effect}, but a complete phase constant function can’t be directly derived from it, hindering the measurement of higher-order dispersion through this appliance. The attempt to recover the phase function of a complex function from its modulus function and extra information is called the phase retrieval problem \cite{J1982Phase}. 

The phase retrieval problem was first discussed in Image Recovery,
and there have been a bunch of algorithms that have been proven to converge mathematically or behave robustly and converge fast in practice, using in subjects like Crystallography and Astronomy \cite{Fienup1987Reconstruction, Gerchberg1972APA, Yudilevich1987RestorationOS}.

Although the phase retrieval problem in a two-photon interference(TPI) model hasn't been discussed before, there is still a clue that either the HOM interference or the phase retrieval problem has connections with the autocorrelation technique for ultra-fast optics: The HOM interference has been used for measuring the time duration of the ultrafast pulse, like the autocorrelation technique \cite{Y1993Measurement};
The phase retrieval algorithms have been adapted and used in the Frequency Resolved Optical Gating (FROG), which was invented for complete pulse characterization and can be seen as an upgrade of the autocorrelation technique \cite{1994Frequency,  1997Measuring}.

In this research, two algorithms that have been applied both in the domain of Image Recovery and FROG were adapted for the TPI-type phase retrieval problem and used to recover the phase constant of the medium in the interferometer mentioned above. One of these algorithms is called the Gerchberg-Saxton(G-S) algorithm, the other algorithm is called the generalized projection(GP) algorithm.
Furthermore, this TPI-type phase retrieval problem can easily be generalized to the recovery of the phase distribution of a single-photon pulse, and even the joint spectrum phase(JSP) of photon pairs, which has few relevant studies only in recent years\cite{2016Phase, 2019Measuring, 2021Measuring, 0Joint}. 

The paper is structured as follows, In Sec. 2, we discuss the relationship between TPI and phase retrieval problem from the HOM effect with independent fields to an unbalanced fourth-order interferometer. In Sec. 3, we introduce the idea and our realization method of these algorithms. In Sec. 4, we provide simulation outcomes to show the performance of the algorithms. In Sec. 5, we discussed the perspective use of the TPI-type phase retrieval problem in the measurement of JSP. In Sec. 6, we conclude.

\section{TPI-type phase retrieval problem}

Considering the HOM interference of independent pulses with the same photon statistic feature, as is shown in Fig. 1(a), two independent pulses with a normalized field function ${{E}_{1}}\left( t \right)$, ${{E}_{2}}\left( t \right)$ in the same polarization incident in a BS simultaneously, and they are detected by the single-photon detectors at the other end. 

The rate of registering a photon by a detector with the time window is${{R}_{A}}$ and ${{R}_{B}}$, and the coincidence rate ${{R}_{C}}$ is the rate of registering a photon at the two detectors at the same time window. A normalized coincidence rate ${{N}_{C}}$ is defined by,
\begin{equation}
{{N}_{C}}\left( \tau  \right)=\frac{{{R}_{C}}\left( \tau  \right)}{{{R}_{C}}\left( \infty  \right)}
\end{equation}
$\tau$ is the relative delay between the incident photons from the different ways and according to the HOM effect, and ${{N}_{C}}$can be deduced in this way, 
\begin{equation}
{{N}_{C}}\left( \tau  \right)=1-\xi V\left( \tau  \right)
\end{equation}
Here $V\left( \tau  \right)$ is called the interference visibility function or the mode matching degree, which is defined by:
\begin{equation}
V\left( \tau  \right)={{\left| \int{dtE_{1}^{*}\left( t \right)E_{2}^{{}}\left( t+\tau  \right)} \right|}^{2}}={{\left| \int{d\omega \tilde{E}_{1}^{*}\left( \omega  \right)\tilde{E}_{2}^{{}}\left( \omega  \right){{e}^{i\omega \tau }}} \right|}^{2}}
\end{equation}
$\xi $ is a coefficient that equals 1 1/2 1/3 respectively when the quantum state of incident fields are both in the single-photon state, coherent state, and thermal state and $V\left( \tau  \right)$ is the square modulus of the cross-correlation function. According to the generalized Wiener-Khintchine theorem, the cross-correlation function can be written as the Fourier transform of the cross-spectrum density function, which is also shown in Eq. (3) \cite{PhysRevLett.59.2044, PhysRevResearch.4.023125, 1987Relation}.

Obviously, $V\left( \tau  \right)$ can be deduced from $\tilde{E}_{1}^{{}}\left( \omega  \right)$ and $\tilde{E}_{2}^{{}}\left( \omega  \right)$, the function of incident fields but not vice versa because of the lost of phase information in square modulus.
As the incident fields spectrums ${{\left| \tilde{E}\left( \omega  \right) \right|}^{2}}$ is measurable, it is hopeful to reconstruct the relative phase function $\exp \left( i{{\varphi }_{2}}\left( \omega  \right)-i{{\varphi }_{1}}\left( \omega  \right) \right)$ of the incident fields 
from interference visibility function $V\left( \tau  \right)$ with their spectrum ${{\left| \tilde{E}\left( \omega  \right) \right|}^{2}}$ as extra information. 
This quest for phase information from a measured squared magnitude function and extra mathematic constraints is called the phase retrieval problem \cite{J1982Phase}.


\begin{figure}[ht]    \centering\includegraphics[width=12cm]{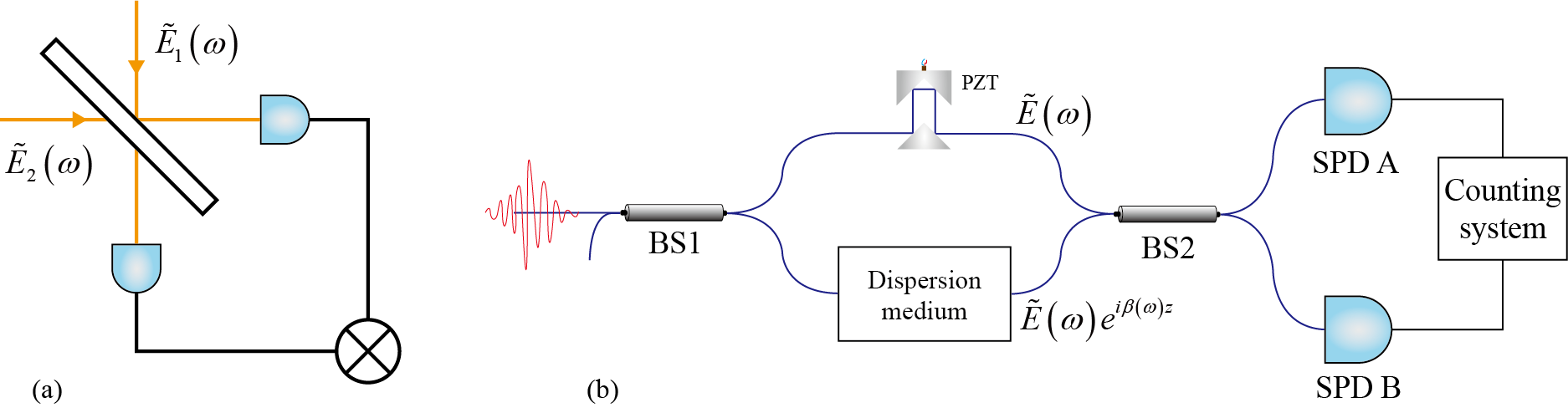}
    \caption{Device illustration. 
    (a) The HOM interferometer.
    (b) The unbalanced fourth-order interferometer.}
\end{figure}

The unbalanced M-Z interferometer for dispersion measurement referend in Sec. 1 is shown in Fig. 1(b). In this appliance, the piezoelectric transducer(PZT) is used to average out the impact of first-order interference. As the incident pulses of the second BS both come from the same light source, a mode-locked laser, and only one of the pulses passed through a medium with dispersion, the intensity spectrum of these pulses will still be the same, and the measured visibility function will be,
\begin{equation}
V\left( \tau  \right)={{\left| \int{d\omega {{{\tilde{E}}}^{*}}\left( \omega  \right)\tilde{E}\left( \omega  \right){{e}^{-i\beta \left( \omega  \right)z}}{{e}^{i\omega \tau }}} \right|}^{2}}={{\left| \int{d\omega I\left( \omega  \right){{e}^{-i\beta \left( \omega  \right)z}}{{e}^{i\omega \tau }}} \right|}^{2}}
\end{equation}
$\beta \left( \omega  \right)$ is the phase constant, the imaginary part of the propagation constant, which specifies the phase change of different frequencies per unit distance.
Generally, the phase constant would be Taylor expanded at a central frequency to discuss the relationship between the visibility function and the component, which is also called the n-order dispersion.
\begin{equation}
\beta \left( \omega  \right)={{\beta }_{0}}+{{\beta }_{1}}\left( \omega -{{\omega }_{0}} \right)+\frac{1}{2}{{\beta }_{2}}{{\left( \omega -{{\omega }_{0}} \right)}^{2}}+...
\end{equation}
Considering their impact on $V\left( \tau  \right)$, the impact of ${{\beta }_{0}}$ is erased by modulus squared operation, and ${{\beta }_{1}}$ is inversely proportional to the group delay in fiber, whose impact is mixed with relative delay $\tau$  in the visibility function. So the phase constant information we are able to get correctly from the visibility function is ${{\beta }_{2}}$, which is also the representation of GVD, and other higher-order dispersion. In other words, the use of the phase retrieval algorithm here is trying to recover $\beta'' \left( \omega  \right)$ correctly.

When the incident field is in the Hermite-Gaussian time mode and only GVD is considered, the GVD and $V\left( \tau  \right)$ can be deduced directly from each other by an analytic relationship \cite{2021Propagation}. GVD is just the second-order component of phase constant, and when we want to measure higher-order coefficients with an incident pulse of an arbitrary spectrum, things get complicated
as the phase constant function $\beta \left( \omega  \right)$ can’t be derived directly from the visibility function $V\left( \tau  \right)$, spectrum $I\left(\omega\right)$ and the medium length $z$, and this is what a phase retrieval algorithm trying to do. 
In other words, the specific phase retrieval problem we worked on in this study is trying to get phase constant function $\beta \left( \omega  \right)$ with the known of $V\left( \tau  \right)$, $I\left( \omega  \right)$ and z, with relation in Eq. (4), and here the spectrum can be seen as extra information exclude the modulus of cross-correlation function $\sqrt{V\left( \tau  \right)}$.

\section{Realization of the two algorithms}
For this TPI-type phase retrieval algorithm,
We transplanted two pulse-retrieval algorithms that had been realized and checked in other types of phase retrieval problems before, for example, image recovery, and FROG (frequency-resolved optical gating).
One of them is the Gerchberg-Saxton algorithm (G-S algorithm), and the other is the generalized projection algorithm (GP algorithm). 
These two algorithms are based on essentially similar principles that call for iterative Fourier transformation back and forth between the two domains and manipulation using data or constraints in these domains between Fourier transforms \cite{J1982Phase, Gerchberg1972APA, 1994Improved}.


The G-S algorithm is one of the most basic algorithms for phase retrieval problems, and it was first presented for reconstructing the phase of the wave function whose intensity in the Fraunhofer diffraction and the imaging planes are known \cite{Gerchberg1972APA}. 
From Eq. (4), $V\left( \tau  \right)$ is the square modulus of Fourier transform of $I\left( \omega  \right)\exp \left( i{{\beta }_{k+1}}\left( \omega  \right)z \right)$, while ${{I}^{2}}\left( \omega  \right)$ is the square modulus of it, similar to the relationship among the intensity functions on the two planes and the wave function.
This transplantation of the algorithm was realized by replacing the amplitude functions in the original algorithm with $\sqrt{V\left( \tau  \right)}$ and the ${I}\left( \omega  \right)$ respectively.


\begin{figure}[ht]
    \centering\includegraphics[width=7cm]{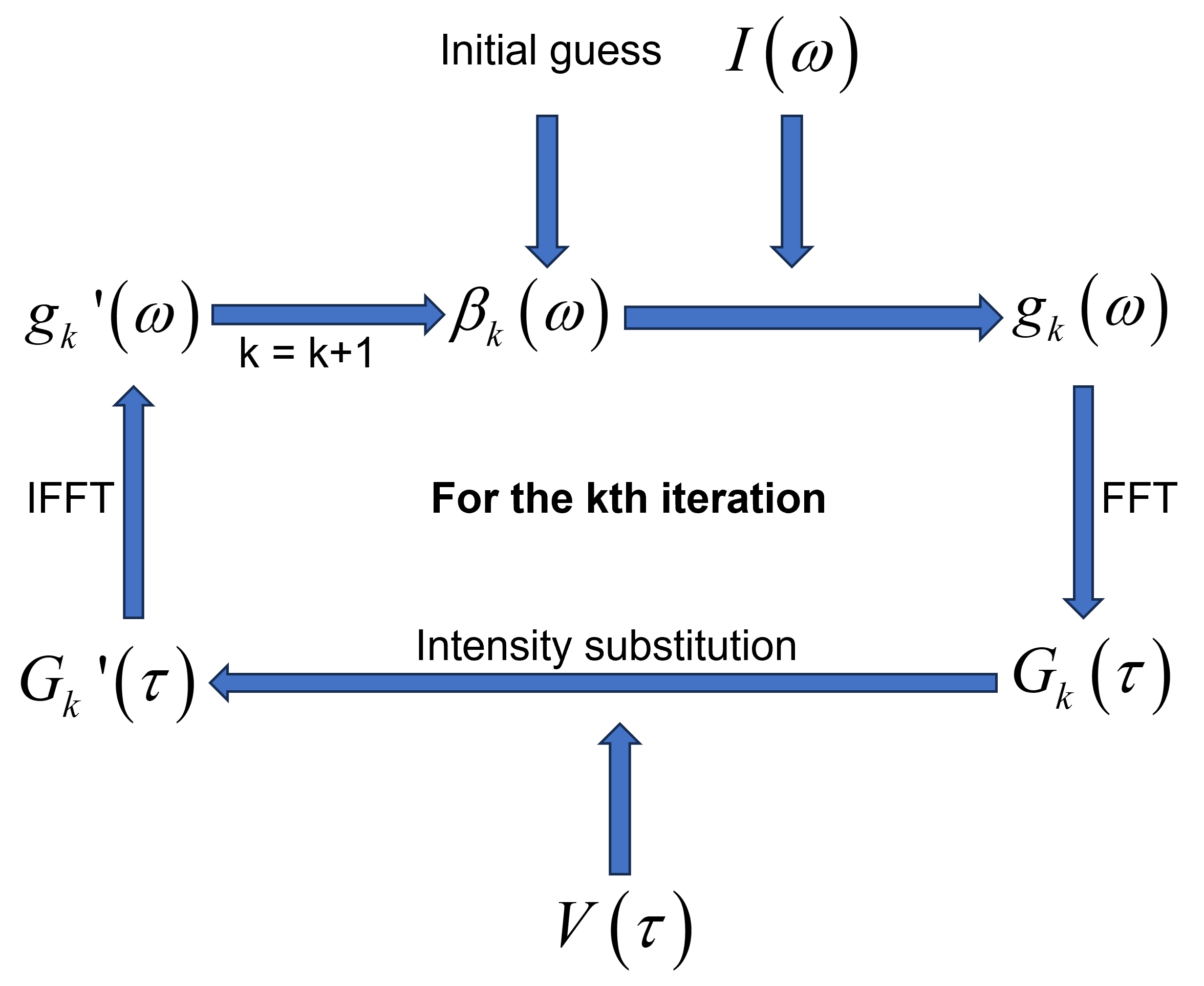}
    \caption{Iteration procedure of the G-S algorithm.}
\end{figure}
The algorithm here can be logically depicted in Fig. 2. ${{g}_{k}}\left( \omega  \right)$ is the current guess of cross-spectrum density of the pulse in the two fields defined by
\begin{equation}
{{g}_{k}}\left( \omega  \right)=I\left( \omega  \right)\exp \left( i{{\beta }_{k}}\left( \omega  \right)z \right)
\end{equation}
and ${{G}_{k}}\left( \tau  \right)$ is the current guess of the cross-correlation function.
The algorithm works like this: Starting with an initial guess of phase constant ${{\beta }_{0}}\left( \omega  \right)$, the cross-spectrum density ${{g}_{0}}\left( \omega  \right)$ can be generated. This function is then Fourier transformed with respect to omega in order to generate the cross-correlation function ${{G}_{0}}\left( \tau  \right)$ in the time domain. The measured $V\left( \tau  \right)$ is then used to generate a new cross-correlation function ${{G}_{0}}'\left( \tau  \right)$ by replacing the magnitude of ${{G}_{0}}\left( \tau  \right)$ with $\sqrt{V\left( \tau  \right)}$, since $\left|  {{G}_{0}}\left( \tau  \right)  \right|$ should get close to $\sqrt{V\left( \tau  \right)}$.
${{G}_{0}}'\left( \tau  \right)$ is then transformed back into the frequency domain by applying an inverse Fourier transform. In the last step of the cycle, the modified cross-spectrum density function ${{g}_{0}}'\left( \omega  \right)$ is used to generate a new guess of phase constant ${{\beta }_{1}}\left( \omega  \right)$ with Eq. (6) and all the subsequent iterations can work following these steps. For the k iteration, this iteration procedure can be written as,
\begin{equation}
{{G}_{k}}\left( \tau  \right)=\int{{{g}_{k}}\left( \omega  \right)\exp \left( -i\omega \tau  \right)d\omega }
\end{equation}

\begin{equation}
{{G}_{k}}'\left( \tau  \right)=\sqrt{V\left( \tau  \right)}\frac{{{G}_{k}}\left( \tau  \right)}{\left| {{G}_{k}}\left( \tau  \right) \right|}
\end{equation}

\begin{equation}
{{g}_{k}}'\left( \omega  \right)=\frac{1}{2\pi }\int{{{G}_{k}}'\left( \tau  \right)\exp \left( i\omega \tau  \right)d\tau }
\end{equation}

\begin{equation}
{{g}_{k+1}}\left( \omega  \right)=I\left( \omega  \right)\frac{{{g}_{k}}'\left( \omega  \right)}{\left| {{g}_{k}}'\left( \omega  \right) \right|}
\end{equation}

A normalized squared error for supervising the convergence of the algorithm is defined by:

\begin{equation}
{{E}_{k}}=\frac{\sum\limits_{i}{{{\left( \left| {{G}_{k}}\left( {{\tau }_{i}} \right) \right|-\sqrt{V\left( {{\tau }_{i}} \right)} \right)}^{2}}}}{\sum\limits_{i}{V\left( {{\tau }_{i}} \right)}}
\end{equation}
The convergence of this algorithm has been proved by showing that the defined error is decreasing or staying at each iteration \cite{Gerchberg1972APA}.

The generalized projection algorithm was first invented to reconstruct an image from the phase-sign function and the intensity function on the Fraunhofer diffraction plane, which are classified as the mathematic constraint and Fourier constraint \cite{1172785}. 
There are two solution sets that satisfy each of these constraints and a wanted solution that satisfies both of the constraints will be at the intersection of the two sets.

The phase-sign function can be generalized to other kinds of mathematical constraints, for example in the FROG-type phase retrieval problems, the mathematical constraint is the relationship given by nonlinear optic equations. For the TPI-type phase retrieval problem, the $\sqrt{V\left( \tau  \right)}$ can be regarded as the Fourier constraint just like what we did in the G-S algorithm and the ${I}\left( \omega  \right)$ can be regarded as the mathematic constraint that rules the amplitude of the magnitude of $I\left( \omega  \right)\exp \left( i{{\beta }_{k}}\left( \omega  \right)z \right)$.


\begin{figure}[ht]    \centering\includegraphics[width=12cm]{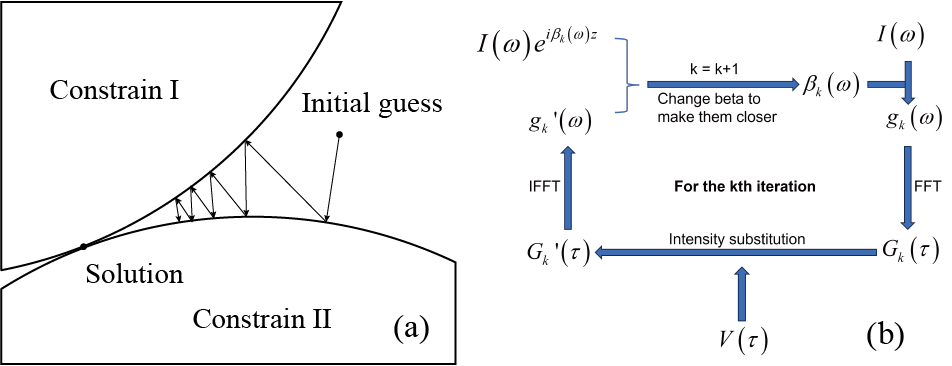}
    \caption{Idea and iteration procedure of the GP algorithm. 
    (a) The idea of projection.
    (b) Iteration procedure of the GP algorithm.}
\end{figure}

The idea of the projection method is to define an initial value function and continuously make projections of the function on two solution sets, which means finding the solution closest to the current function in the solution set as the next generation functions as shown in Fig. 3(a). When the two constraints are convex, convergence is guaranteed, and the generalized projection is defined when the two constraints are not convex. The GP algorithm is robust when applied to FROG and image recovery, so it's worthwhile to realize this algorithm in the TPI-type phase retrieval problem. The algorithm here can be logically depicted in Fig. 3(b) \cite{Yudilevich1987RestorationOS, Delong1994PulseRI, youla1982image}
.

In this problem, the projection that satisfies the visibility constraint means substituting the magnitude of  ${{G}_{k}}\left( \tau  \right)$ with the measured data, which is the same as the operation implemented in the G-S algorithm. The projection that satisfies the spectrum constraint comes differently, as it wants a ${{g}_{k+1}}\left( \omega  \right)$ that satisfies the spectrum constraint and minimizes a distance Z, which is defined by

\begin{equation}
Z=\sum\limits_{i}{{{\left( I\left( {{\omega }_{i}} \right){{e}^{i{{\beta }_{k+1}}\left( {{\omega }_{i}} \right)z}}-{{g}_{k}}'\left( {{\omega }_{i}} \right) \right)}^{2}}}
\end{equation}

Z depicts the difference between ${{g}_{k+1}}\left( {{\omega }_{i}} \right)$ and ${{g}_{k}}'\left( {{\omega }_{i}} \right)$whose phase factor are ${{\varphi }_{k+1}}\left( {{\omega }_{i}} \right)$and ${{\varphi }_{k}}'\left( {{\omega }_{i}} \right)$ respectively, and Z can be seen as a function of  $\left\{ {{\varphi }_{k+1}}\left( {{\omega }_{i}} \right) \right\}$, which is the phase factor of ${{g}_{k+1}}\left( {{\omega }_{i}} \right)$ at all the sampling frequency $\left\{ {{\omega }_{i}} \right\}$, so we can make a gradient descent to find $\left\{ {{\varphi }_{k+1}}\left( {{\omega }_{i}} \right) \right\}$ that minimize Z. And piratically, it's not necessary to find the minimum of Z on each iteration, as finding a one-dimensional minimization along the opposite direction of the gradient starting at the point of the last iteration is sufficient for convergence. The reason is that the other steps will make up for this inaccurate projection.
The one-dimensional minimization step was realized by an adaptive learning rate gradient descent algorithm.
The partial derivates of Z with respect to $\left\{ {{\varphi }_{k+1}}\left( {{\omega }_{i}} \right) \right\}$ are,
\begin{equation}
{{\left. \frac{\partial Z}{\partial {{\varphi }_{k+1}}\left( {{\omega }_{i}} \right)} \right|}_{{{\varphi }_{k+1}}\left( {{\omega }_{i}} \right)={{\varphi }_{k}}\left( {{\omega }_{i}} \right)}}=2\left| {{g}_{k}}'\left( {{\omega }_{i}} \right) \right|I\left( {{\omega }_{i}} \right)\sin \left( {{\varphi }_{k+1}}\left( {{\omega }_{i}} \right)-{{\varphi }_{k}}'\left( {{\omega }_{i}} \right) \right)
\end{equation}

Also, the beta coefficients can be regarded as the parameters of Z to implement one-dimensional gradient descent. Considering the lower-order component only is more intuitive when measurement focuses on the lower-order component, and it converges faster if we only want to estimate the GVD of an optical medium. And the partial derivates of Z with respect to $\left\{ {{\beta }_{j}} \right\}$ are,
\begin{equation}
{{\left. \frac{\partial Z}{\partial {{\beta }_{k+1,j}}} \right|}_{{{\beta }_{k+1,j}}={{\beta }_{k,j}}}}=\frac{{{\sum\limits_{i}{2\left| {{g}_{k}}'\left( {{\omega }_{i}} \right) \right|I\left( {{\omega }_{i}} \right)\sin \left( {{\varphi }_{k+1}}\left( {{\omega }_{i}} \right)-{{\varphi }_{k}}'\left( {{\omega }_{i}} \right) \right)\left( {{\omega }_{i}}-{{\omega }_{0}} \right)}}^{j}}}{\sum\limits_{i}{4I\left( {{\omega }_{i}} \right)}}
\end{equation}

The algorithms we transplanted are two of the most basic phase retrieval algorithms, and they can easily be modified into algorithms like short-cut algorithms and over-correction algorithms with simple adjustments like research in Ref \cite{1994Improved, Krumbuegel2000}.

Iteration procedures of these algorithms mentioned above are interconnected, making it possible to combine those basic iterative algorithms together by switching them between two iterations as a composite algorithm.
In other domains of phase retrieval problems, the composite usually shows better convergence, robustness, and less sensitivity to the initial guess, indicating that it is important to realize multiple algorithms based on different ideas. The combination of algorithms is usually adopted for commercial use, for example, the Femotosoft and MarkTech \cite{1994Improved, Krumbuegel2000, 2003Composite}. A similar attempt was tested in this research.

\section{Simulation}
Simulations were designed to verify the convergence of the algorithms.
First, we set the pre-set phase constant $\beta\left( \omega  \right)$ to be recovered, a spectrum function of the incident beam $I\left( \omega  \right)$ and the length of fiber z, so that the visibility function could be derived using Eq. (4) and used as known data together with the spectrum in phase retrieval algorithm, and the guess of the last iteration would be the result of this algorithm.

Comparing the current guess of the square modulus of the cross-correlation function $\left|G_{k} \left(\tau\right)\right|^2$and the second order derivation of the phase constant $\beta_{k}'' \left(\omega\right)$ with pre-set data, we can check whether the result comes to a reasonable solution.
Convergence is described by the error defined by Eq. (11) changing with the number of iterations.
For the simulation below, the spectrum is in Gaussian shape with a width of $1 nm$ and the center of the spectrum is $1533 nm$. Dispersion medium is a $3.7 km$ long fiber with ${{\beta }_{2}}=4p{{s}^{2}}/km$ and ${{\beta }_{3}}=0.06p{{s}^{3}}/km$. The sampling interval of the spectrum is $0.002 THz$, and the sampling interval of relative time delay is $1/3 ps$, which corresponds to a $0.05 mm$ scanning step size of a right angle prism delay line, and all of these measurement parameters can be realized in practice. 


\begin{figure}[ht]
    \centering\includegraphics[width=12cm]{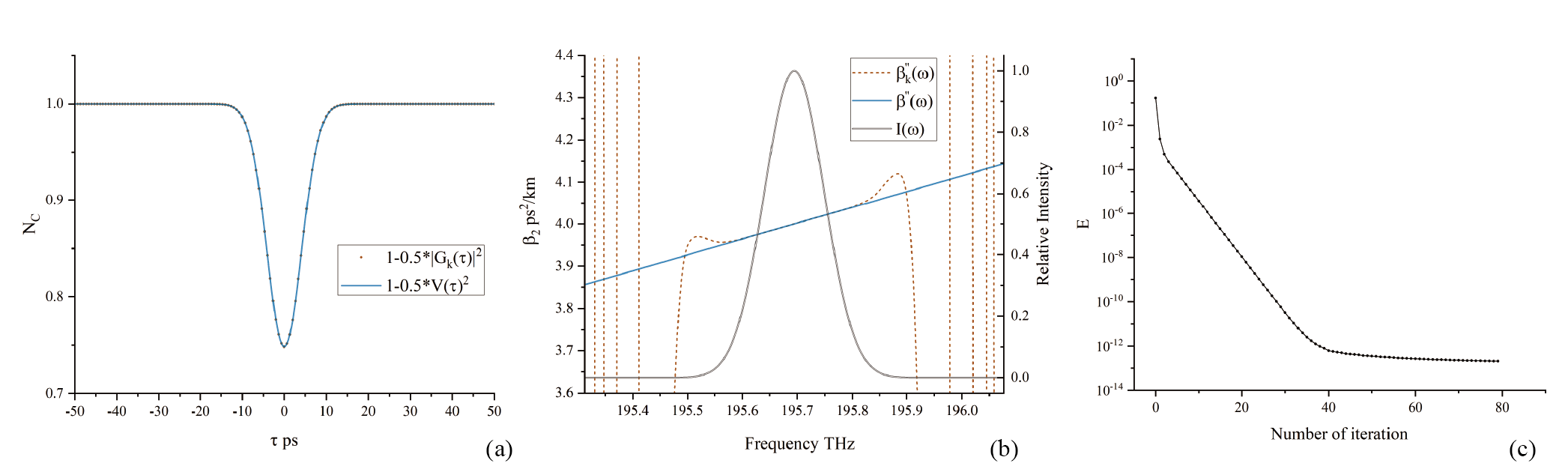}
    \caption{Simulation for the Gerchberg-Saxton algorithm. 
    (a) Comparison between $\left|G_{k} \left(\tau\right)\right|^2$ in dotted red line and $V\left( \tau \right)$ in solid blue line.
    (b) Comparison between $\beta'' \left(\omega\right)$ in dotted red line and $\beta_{k}'' \left(\omega\right)$ in solid blue line, with $I \left(\omega \right) $ as reference spectrum.
    (c) A convergence curve starts from the initial guess to the 80th iteration.}
\end{figure}
In Fig. 4(a) and Fig. 4(b), the solid blue line represents the normalized coincidence function and phase constant function corresponding to pre-set data, dotted red lines and dots represent the corresponding guess in the last iteration, and double solid black lines represent the spectrum of the optic source. Comparing the guess of the last generation with pre-set data can be used to indicate whether the algorithm converges to a desired result. 
In Fig. 4(b), $\beta_{k}'' \left(\omega\right)$ is accurate at the mid-frequency of the spectrum, but this accuracy will decrease as the intensity decreases away from the central frequency. This relation is reasonable because the frequency with smaller intensity contributes less to the interference pattern, and in return, the phase information given by the interference pattern at these frequencies will be more inaccurate \cite{Delong1994PulseRI}. At the frequency with no intensity, the recovered phase is a random value.
Fig. 4(c) shows a convergence curve, and the error decreases to $lg-29.2148$ in the 80th iteration.


\begin{figure}[ht]    \centering\includegraphics[width=12cm]{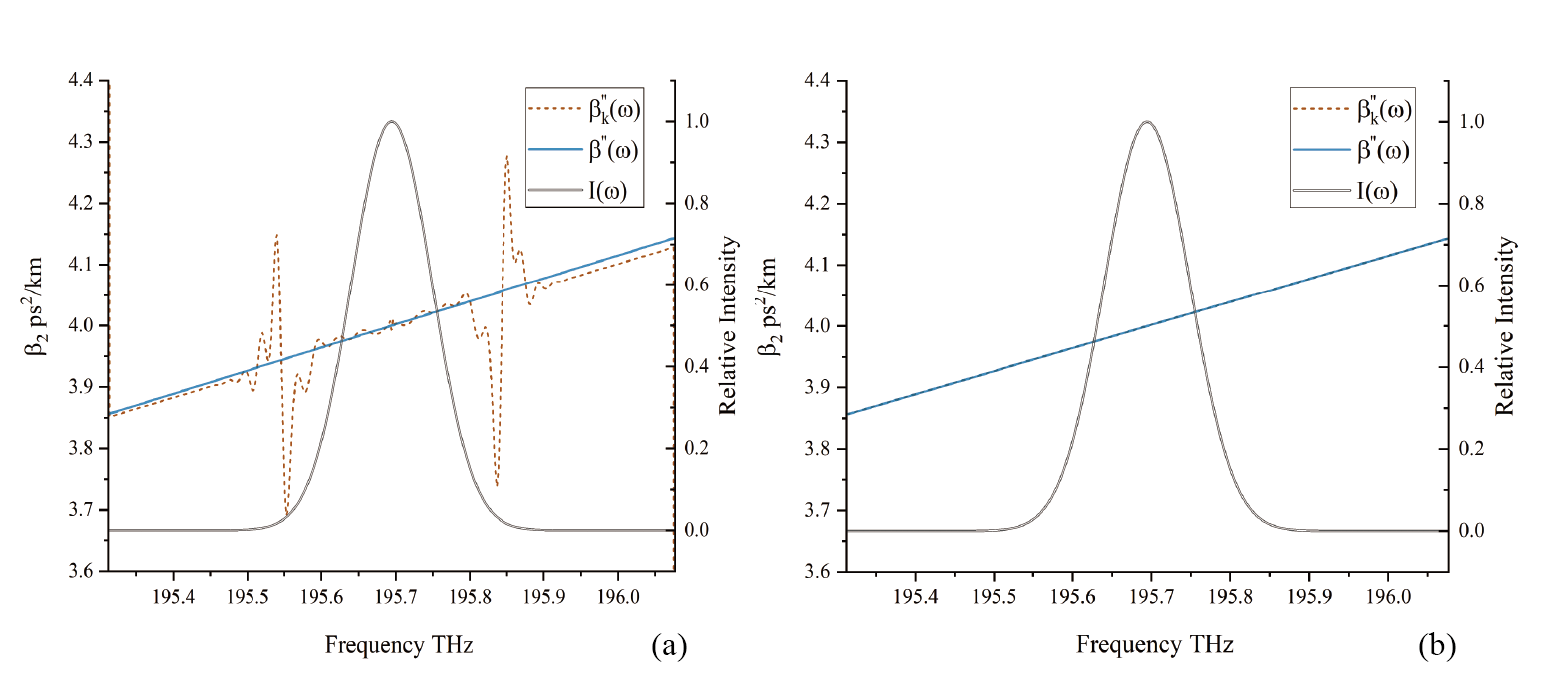}
    \caption{Simulation for the generalized projection algorithm. 
    (a) Comparison between $\beta'' \left(\omega\right)$ in dotted red line and $\beta_{k}'' \left(\omega\right)$ in solid blue line, with $I \left(\omega \right) $ as reference spectrum when regarding $\left\{ \varphi \left( {{\omega }_{i}} \right) \right\}$ as variables.
    (b) Comparison between $\beta'' \left(\omega\right)$ in dotted red line and $\beta_{k}'' \left(\omega\right)$ in solid blue line, with $I \left(\omega \right) $ as reference spectrum when regarding $\left\{ {{\beta }_{j}} \right\}$ as variables.}
\end{figure}

Fig. 5 (a)(b) are the outcomes of GP algorithm that regards $\left\{ \varphi \left( {{\omega }_{i}} \right) \right\}$ and $\left\{ {{\beta }_{j}} \right\}$ as variables to implement gradient descent respectively(the GP1 algorithm and GP2 algorithm). The outcome showed in Fig. 5 (a) is rougher than that of the G-S algorithm while the $\beta_{k}''\left( \omega  \right)$ in Fig. 5 (b) fits $\beta''\left( \omega  \right)$ perfectly.

The simulation outcome of the composite algorithm introduced in Sec. 3 is shown in Fig. 6, and more data about these simulations are summarized in Table 1.


\begin{figure}[ht]    \centering\includegraphics[width=12cm]{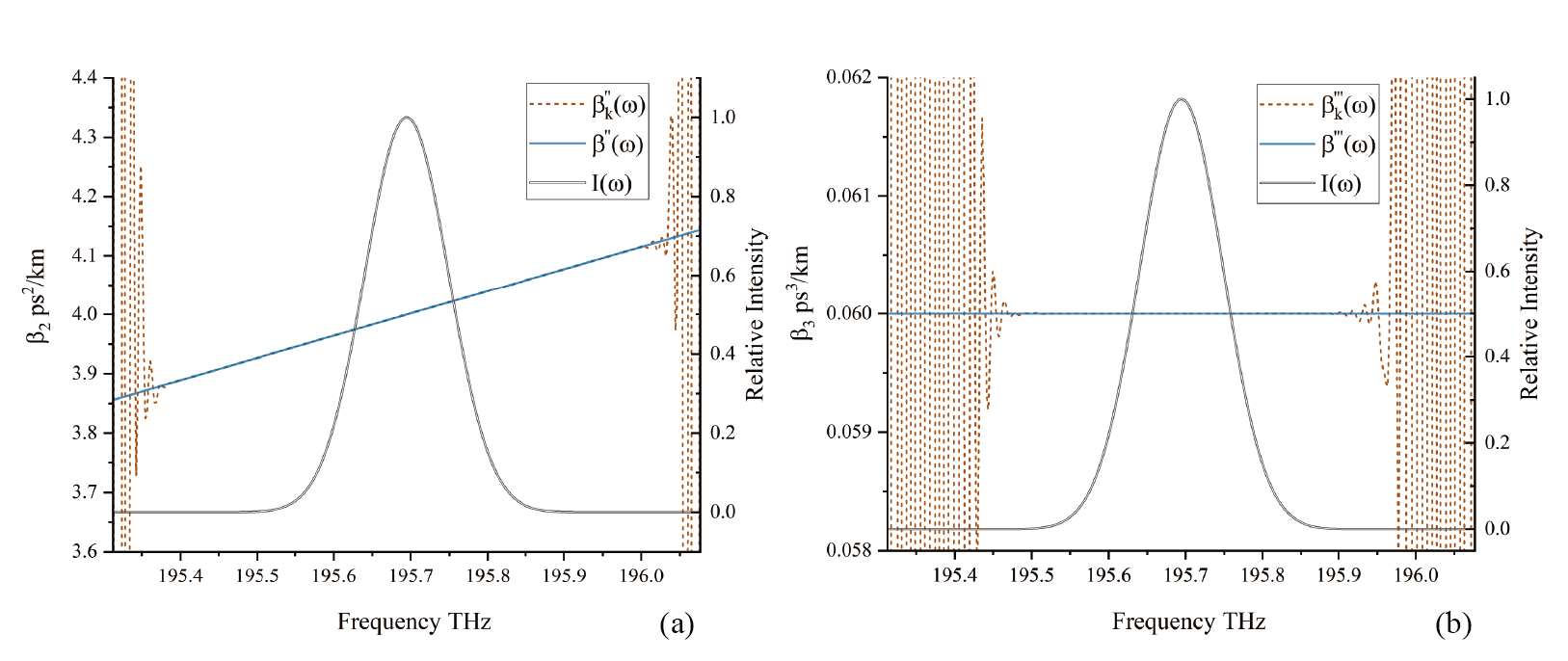}
    \caption{Simulation for the composite algorithm. 
    (a) Comparison between $\beta'' \left(\omega\right)$ in dotted red line and $\beta_{k}'' \left(\omega\right)$ in solid blue line, with $I \left(\omega \right) $ as reference spectrum.
    (b) Comparison between $\beta''' \left(\omega\right)$ in dotted red line and $\beta_{k}''' \left(\omega\right)$ in solid blue line, with $I \left(\omega \right) $ as reference spectrum.}
\end{figure}

\begin{table}[htp]
\begin{center}
\textbf{Table 1}~~Results of simulations.\\
\end{center}
\resizebox{\textwidth}{15mm}{
\begin{tabular}{@{}llllll@{}}
\toprule
     & $\beta_2 $ error ($p{s}^{2}/km$) & $\beta_3 $ error ($p{s}^{3}/km$) & Number of iteration & E of the initial guess & E of the last iteration \\ \midrule
G-S       & 8.08564E-06 & 6.87183E-05 & 80                  & 0.16641966                 & 2.05E-13                    \\
GP1       & 6.84294E-05 & 0.005335055 & 1897                & 7.50E-09                   & 1.57E-09                    \\
GP2       & 3.56532E-09 & 0.00040857  & 300                 & 0.16641966                 & 2.85E-13                    \\
CMP & 7.43358E-10 & 2.92975E-08 & 94                  & 0.16641966                 & 4.82E-18                    \\ \bottomrule
\end{tabular}}
\end{table}

This composite algorithm took 94 iterations to converge to a more accurate solution with a rougher initial value, behaving better than all the algorithms used alone. 
Taking a third derivative of the phase constant to get $\beta_{k}'''\left( \omega  \right)$, we can see that the value of this function also falls perfectly on $0.06p{{s}^{2}}/km$ as Fig. 6(b), proving that this algorithm is able to recover the high order dispersion even when it has a much smaller effect on the phase constant than the lower-order dispersion.

${{\beta }_{2}}$ and ${{\beta }_{3}}$ in this table are deduced from the $\beta \left( \omega  \right)$ by doing weighted average to $\beta''\left( \omega  \right)$ and $\beta'''\left( \omega  \right)$. as is shown in the table, even the most inaccurate algorithm, the GP1 algorithm has a result with error less than 0.05\%.
The error of the G-S algorithm has been mathematically proven to exist in Ref. \cite{Gerchberg1972APA} and for the GP algorithm the error is induced by the imprecise projection from the Fourier domain to the time domain. Although a single algorithm can converge to a very accurate result with some special initial guess, these examples are not universal and the composite algorithm should be the best choice. The resulting error of the composite algorithm is small enough to be ignored in practice.

\begin{figure}[htpb]
    \centering\includegraphics[width=12cm]{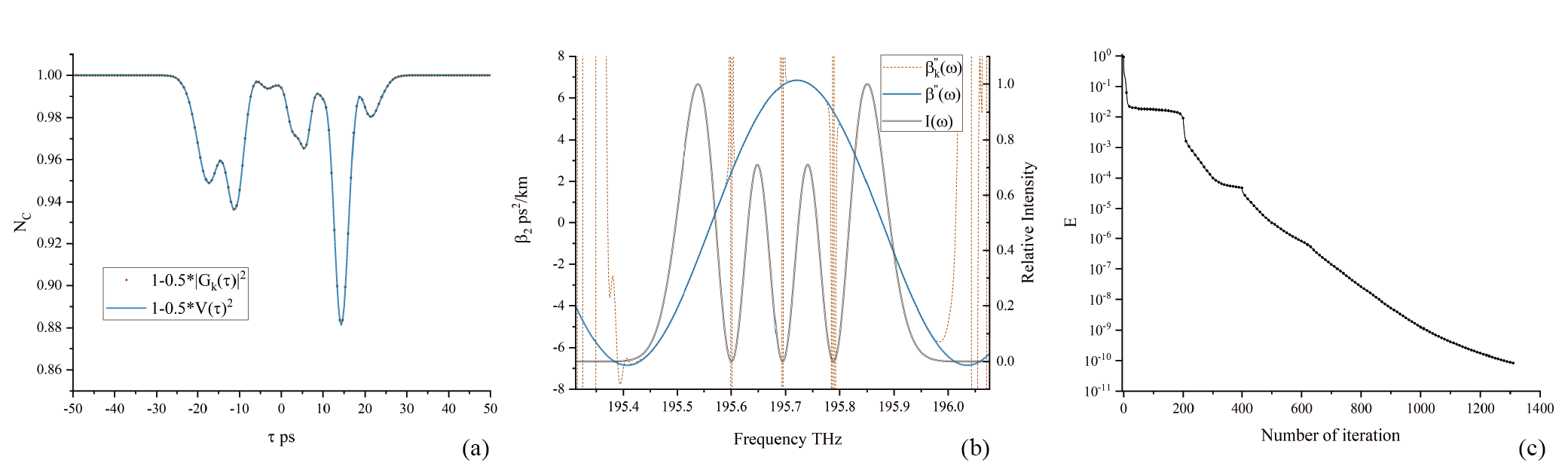}
    \caption{Simulation with optical source in 3-order H-G time mode and $\beta \left(\omega\right)$ in a cosine form. 
    (a) Comparison between $\left|G_{k} \left(\tau\right)\right|^2$ in dotted red line and $V\left( \tau \right)$ in solid blue line.
    (b) Comparison between $\beta'' \left(\omega\right)$ in dotted red line and $\beta_{k}'' \left(\omega\right)$ in solid blue line, with $I \left(\omega \right) $ as reference spectrum.
    (c) A convergence curve starts from the initial guess to the 1310th iteration.}
\end{figure}

The last simulation was used to check if this algorithm is robust enough. The spectrum of the incident beam was changed into third-order Hermitage Gaussian time mode, and the phase constant was set to a cosine function, inducing a much more complex interference figure, while the measurement parameters kept the same.
As is shown in Fig. 7, the algorithm still converges, taking 1310 iterations to reduce the error from 0.929 to 8.235E-11, meaning that the algorithm we have realized is robust enough to be used practically in measurement like Ref. \cite{2021Propagation}.

\section{JSP measurment}

Looking back at Eq. (3), if the magnitude  $\left| {{{\tilde{E}}}_{1}}\left( \omega  \right) \right|$ and $\left| {{{\tilde{E}}}_{2}}\left( \omega  \right) \right|$ are known, the phase difference function can be recovered from $V\left( \tau  \right)$, which means that the phase of an arbitrary pulse whose spectrum intensity is known can be recovered from the HOM dip when the spectral mode of the other incident pulse is known, as a reference pulse.
\begin{figure}[ht]
    \centering\includegraphics[width=10cm]{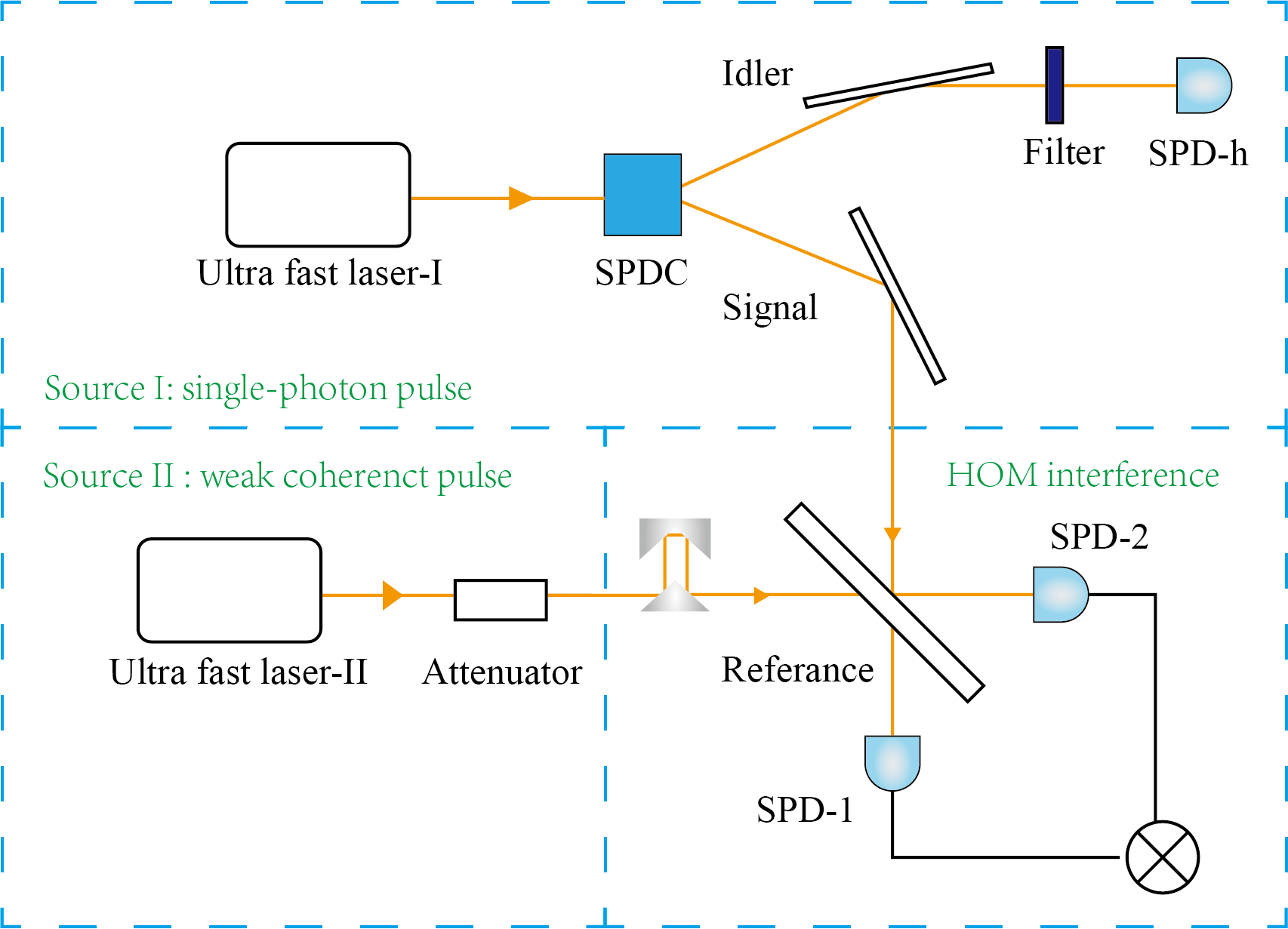}
    \caption{Schematic diagram of JSP measuring apparatus. BS, 50:50 beam splitter;SPD, single-photon detector; SPDC, Spontaneous Parametric Down Conversion}
\end{figure}

A possible measurement of JSP is shown in Fig. 8, the two incident fields are a weak coherent pulse and a transform-limited single-photon pulse generated by gated photon detection. The ultra-fast laser-I pumped a spontaneous parametric down conversion(SPDC) process, generating a photon pair with in $\left| {{0}_{s}},{{0}_{i}} \right\rangle +\xi \int{d{{\omega }_{s}}d{{\omega }_{i}}\Phi \left( {{\omega }_{s}},{{\omega }_{i}} \right){{a}_{s}}^{\dagger }\left( {{\omega }_{s}} \right){{a}_{i}}^{\dagger }\left( {{\omega }_{i}} \right)\left| vac \right\rangle }$. The idler photon will pass through a narrow band filter and arrive at the SPD-h, heralding a single-photon pulse in the direction of the signal photon. The ultra-fast laser-II is synchronized with the ultra-fast laser I and attenuated by an attenuator thus the intensity of the coherent pulse is comparable with the single-photon pulse.
For the coherent pulse source, we have
\begin{equation}
    \widehat{E}\left( t \right)\left| \alpha  \right\rangle =\int{d\omega \alpha \left( \omega  \right){{e}^{i\omega \left( {{t}_{\alpha }}-\tau -t \right)}}\left| \alpha  \right\rangle }=A\left( t \right)\left| \alpha  \right\rangle
\end{equation}
The coherent pulse is peaked at ${{t}_{\alpha }}$ and $\tau $ is introduced by the optics delay line.
For the single-photon pulse source, we have
\begin{equation}
    \widehat{E}\left( t \right)\left| {{\Psi }_{1}} \right\rangle =\int{d\omega K'{{F}_{i}}\left( {{t}_{i}} \right)\tilde{\Phi }\left( \omega ,{{\omega }_{i0}} \right){{e}^{i\omega \left( {{t}_{p}}-t \right)}}\left| vac \right\rangle }={{\Psi }_{1}}\left( t \right)\left| vac \right\rangle 
\end{equation}
where ${{F}_{i}}\left( {{t}_{i}} \right)=\int{d\omega }{{f}_{i}}\left( \omega  \right){{e}^{i\omega \left( {{t}_{i}}-{{t}_{p}} \right)}}$, ${{t}_{i}}$ is the time when the idler photon is detected,  ${{f}_{i}}\left( \omega  \right)$ is the function of the filter, and ${{\omega }_{i0}}$ is center wavelength of the filter.
Under this “narrow band filter” approximation,  ${{\Psi }_{1}}\left( t \right)$ is independent to ${{t}_{i}}$, and it is a normalized function, as $\left| {{\Psi }_{1}} \right\rangle $ is a single-photon pulse. ${{t}_{i}}$ and the width of the time window of the SPD-h on the idler arm only affect the occurrence frequency of the single-photon issue.

To make the following formulas simpler, we set:
\begin{equation}
    \alpha \left( \omega  \right)=A\alpha '\left( \omega  \right)
\end{equation}
\begin{equation}
    K'{{F}_{i}}\left( {{t}_{i}} \right)\tilde{\Phi }\left( \omega ,{{\omega }_{i0}} \right)=\tilde{\Phi }'\left( \omega ,{{\omega }_{i0}} \right)
\end{equation}
where $\alpha '\left( \omega  \right) $ and $\tilde{\Phi }'\left( \omega ,{{\omega }_{i0}} \right) $are both normalized functions.

Assuming that the coupler ration of BS is 50:50, the coincidence rate for the other two detectors at ${{t}_{1}}$ and ${{t}_{2}}$ respectively should be,
\begin{equation}
Rc\left( {{t}_{1}},{{t}_{2}} \right)
=\frac{1}{4}{
{
\left| 
\left( {{\widehat{E}}_{1in}}\left( {{t}_{1}} \right)+{{\widehat{E}}_{2in}}\left( {{t}_{1}} \right)
\right)
\left( {{\widehat{E}}_{1in}}\left( {{t}_{2}} \right)-{{\widehat{E}}_{2in}}\left( {{t}_{2}} \right)
\right)
\left| {{\Psi }_{1}} \right\rangle \left| \alpha  \right\rangle  \right|}^{2}}
\end{equation}
Integrate $Rc$ over ${{t}_{1}}$ and ${{t}_{2}}$ as these two detectors in the HOM interferometer are slower than the pulse, and use the generalized Wiener-Khintchine theorem,
\begin{equation}
    Rc\left( \tau  \right)=\frac{1}{4}\left( {{\left| A \right|}^{4}}+2{{\left| A \right|}^{2}}\int{dt{{\left| {{\Psi }_{1}}\left( t \right) \right|}^{2}}-2{{\left| A \right|}^{2}}{{\left| \int{d\omega \alpha '\left( \omega  \right)\Phi '\left( \omega ,{{\omega }_{i0}} \right)}{{e}^{-i\omega \tau }} \right|}^{2}}} \right)
\end{equation}
${{t}_{\alpha }}-{{t}_{p}}$ is ignored here as it can be covered by the change of $\tau $.
And the normalized coincidence rate would be:
\begin{equation}
    Nc\left( \tau  \right)=1-2\frac{{{\left| \int{d\omega \alpha '\left( \omega  \right)\Phi '\left( \omega ,{{\omega }_{i0}} \right)}{{e}^{-i\omega \tau }} \right|}^{2}}}{{{\left| A \right|}^{2}}+2}
\end{equation}
The visibility $V\left( \tau  \right)=\frac{2{{\left| \int{d\omega \alpha '\left( \omega  \right)\Phi '\left( \omega,{{\omega }_{i0}} \right)}{{e}^{-i\omega \tau }} \right|}^{2}}}{{{\left| A \right|}^{2}}+2}$ is related with the intensity of the coherent state pulse. When ${{\left| A \right|}^{2}}\ll 1$, the interference visibility is close to $100\%$, and visibility will decrease as ${{\left| A \right|}^{2}}$ increases. 
 As this phase retrieval process is insensitive to the overall size of the function, recovering the phase of $\Phi '\left( \omega,{{\omega }_{i0}} \right)$ from $V\left( \tau  \right) $ is the same to the phase recover problem mentioned at the beginning of this part. And the two-dimensional JSP can be obtained by repeating the above-mentioned process at different ${{\omega }_{i0}}$.

The way to measure the JSP is similar to Ref. \cite{2021Measuring}, which also requires a heralding detection and HOM interference, and the idea in this paper may be faster as it only requires a one-dimension data measurement for the HOM interference while the other one requires a two-dimension measurement. However, the JSI is unable to be recovered in this measurement, so a full characterization of the photon-pair still needs extra steps, increasing the complexity of the measurement.

\section{Conclusion}
In this research, we discussed the phase retrieval problem in the two-photon interference model for the first time, and successfully transplanted two algorithms, the G-S algorithm and the GP algorithm into this TPI-type phase retrieval problem as a mathematical process to recover the complete phase constant of the medium from a fourth-order interference pattern and spectrum of optic source.

In the simulation, we verified the convergence and accuracy of the algorithm, the performance superiority of the composite algorithm over algorithms used alone, and the robustness of our algorithm. 
Our algorithm can accurately obtain ${{\beta }_{3}}$ with an error of $lg-7.5332p{{s}^{3}}/km$ and ${{\beta }_{2}}$ with an error of $lg-9.1288p{{s}^{2}}/km$ when ${{\beta }_{3}}=-0.06p{{s}^{3}}/km$, ${{\beta }_{2}}=-4p{{s}^{2}}/km$ and the spectrum of incident field is in Gaussian shape and can converge eve when the incident field is in a three-order Hermitian Gaussian mode and the phase constant is in a cosine shape with a defined error of $8.235E-11$. The result of the simulations shows that this algorithm is competent for recovering the phase constant in real experiments.

For the perspective application of the TPI-type phase retrieval problem, We provided a theoretically possible method to measure the JSP of photon-pair from a spontaneous nonlinear process based on the phase retrieval algorithm. As it requires extra steps of measuring the JSI, this scheme is not convenient enough in comparison with other schemes. Further improvement of this method could focus on utilizing the stimulated-emission-based technique\cite{0Joint} which can increase the intensity of the signal to be measured or multi-photon interference without reference pulse.

For quantum optics, our research provides an algorithmic tool for high-order dispersion measurement using two-photon interference and paves the way for a higher resolution and phase-sensitive quantum tomography. For the phase retrieval problem, our research finds a new application area for the phase retrieval algorithm.
\begin{backmatter}
    \bmsection{Acknowledgments}
    We would like to acknowledge Professor Z.Y. Jeff Ou for constructive discussions.
\end{backmatter}
\bibliography{reference}
\end{document}